\documentstyle[aps,twocolumn,epsfig]{revtex}

\begin{document}
\draft

\twocolumn[\hsize\textwidth\columnwidth\hsize\csname@twocolumnfalse\endcsname

\title{Spin squeezing in nonlinear spin coherent states}
\author{Xiaoguang Wang}
\address{Institute of Physics and Astronomy, Aarhus University, DK-8000, Denmark}
\date{\today}
\maketitle

\begin{abstract}
We introduce the nonlinear spin coherent state via its 
ladder operator formalism and propose a type of nonlinear 
spin coherent state by the nonlinear time evolution of 
spin coherent states. By a new version of
spectroscopic squeezing criteria we study the spin squeezing in both 
the spin coherent state and 
nonlinear spin coherent state. 
The results show that  
the spin coherent state is not squeezed in the $x$, $y$, and $z$ directions, 
and the nonlinear spin coherent state may be 
squeezed in the $x$ and $y$ directions.
\end{abstract}
\pacs{Keywords: Spin squeezing, nonlinear spin coherent states}
]

\section{Introduction}

In nonlinear systems such as optical Kerr medium\cite{Kerr}, squeezed states
of the radiation field\cite{Squeezed} have been extensively studied. The spin
squeezed states are also studied in nonlinear systems\cite{Kitagawa} and
proved to be useful to enhance spectroscopic resolution\cite{Wineland94}, 
{\it e.g.}, in atomic clocks. The generation of the spin squeezed states has
been studied by many ways, such as by interaction of atoms with squeezed
light\cite{Wineland94,A,Kuzmich1,Lukin,Vernac}, quantum nondemolition
measurement of atomic spin states\cite{Kuzmichqnd}, and atomic collisional
interactions\cite{Soerensen}.

The spin squeezed states can be characterized in many different ways\cite
{Kitagawa,Wineland94,Trifonov}. In our study we employ the criteria of spin
squeezing recently proposed by S\o rensen {\it et al.}\cite{Nature} as a new
version of spectroscopic squeezing \cite{Wineland94,Agarwal}. The squeezing
parameter is defined as\cite{Nature} 
\begin{equation}
\xi _{\vec{n}_1}^2={\frac{2j(\Delta J_{\vec{n}_1})^2}{\langle J_{\vec{n}%
_2}\rangle ^2+\langle J_{\vec{n}_3}\rangle ^2}},  \label{eq:xi}
\end{equation}
where $J_{\vec{n}}=\vec{n}\cdot \vec{J},$ $\vec{n}_i(i=1,2,3)$ are
orthogonal unit vectors and $\vec{J}$ is the spin-$j$ angular momentum operator. The
states with $\xi _{\vec{n}}^2<1$ are spin squeezed in the direction $\vec{n}$. 
We will show a
interesting feature that the squeezing parameters $\xi _x^2=\xi _y^2=\xi
_z^2=1$ for spin coherent states(SCS)\cite{SCS}, which indicates that the
SCS is not squeezed in the $x$, $y$ and $z$ directions.

In this paper we introduce the nonlinear SCS and consider the spin squeezing in it. 
Sec. II gives the definition of the nonlinear SCS via its ladder operator
formalism and propose a type of nonlinear SCS by the time evolution of the
SCS under nonlinear Hamiltonian. In Sec. III we first prove that the SCS
exhibits no squeezing in the $x$, $y$, and $z$ directions and then study the
spin squeezing in the nonlinear SCS. A conclusion is given in Sec. IV.

\section{Nonlinear SCS}

We work in a $(2j+1)$-dimensional angular momentum Hilbert space $%
\{|j,m\rangle ;m=-j,...,+j\}$. It is convenient to define a number operator $%
{\cal N}=J_z+j$, and the `number states' $|n\rangle $ which satisfy 
\begin{equation}
|n\rangle \equiv |j,-j+n\rangle ,{\cal N}|n\rangle =n|n\rangle .
\end{equation}

The SCS is given by\cite{SCS}, 
\begin{equation}
|\eta \rangle =(1+|\eta |^2)^{-j}\sum_{n=0}^{2j}{{%
{2j \choose n}%
}}^{1/2}\eta ^n|n\rangle ,  \label{eq:scs}
\end{equation}
where $\eta $ is complex.

It is easy to check that the SCS satisfies the following equation 
\begin{equation}
J_{-}|\eta \rangle =\eta (2j-{\cal N})|\eta \rangle .  \label{laddercs1}
\end{equation}
where the operators $J_{\pm }=J_x\pm iJ_y$. This is a ladder operator
formalism of the SCS.

By recalling the definition of the bosonic nonlinear coherent state\cite
{NLCS} and su(1,1) nonlinear coherent states\cite{SU11}, it is natural to
define the nonlinear SCS as 
\begin{equation}
f({\cal N})J_{-}|\eta \rangle _{nl}=\eta (2j-{\cal N})|\eta \rangle _{nl},
\label{laddernlcs}
\end{equation}
where $f({\cal N})$ is a nonlinear function of the number operator ${\cal N}$%
. Eq.(\ref{laddernlcs}) is of the ladder operator form.

Now we propose a type of nonlinear SCS. The SCS under the evolution of the
nonlinear Hamiltonian $F({\cal N})$ is directly given by 
\begin{eqnarray}
|\eta ,t\rangle  &=&e^{-itF({\cal N})}|\eta \rangle   \nonumber \\
&=&(1+|\eta |^2)^{-j}\sum_{n=0}^{2j}{{%
{2j \choose n}%
}}^{1/2}\eta ^ne^{-itF(n)}|n\rangle .  \label{eq:nl}
\end{eqnarray}
From Eqs.(\ref{laddercs1}) and (\ref{eq:nl}) we find the state $|\eta
,t\rangle $ satisfies 
\begin{equation}
e^{it[F({\cal N}+1)-F({\cal N})]}J_{-}|\eta ,t\rangle =\eta (2j-{\cal N}%
)|\eta ,t\rangle .  \label{eq:kerr}
\end{equation}
According to the definition of the nonlinear SCS, the above state is a
nonlinear SCS with the nonlinear function $e^{it[F({\cal N}+1)-F({\cal N})]}$%
.

For $F({\cal N})={\cal N}^2-{\cal N}$, Eq.(\ref{eq:kerr}) reduces to 
\begin{equation}
e^{i2{\cal N}t}J_{-}|\eta ,t\rangle =\eta (2j-{\cal N})|\eta ,t\rangle .
\label{eq:free2}
\end{equation}
Further let $t=\pi /2,$ the above equation reduces to 

\begin{equation}
\Pi J_{-}|\eta ,\pi /2\rangle =\eta (2j-{\cal N})|\eta ,\pi /2\rangle ,
\end{equation}
where $\Pi =(-1)^{{\cal N}}$ is the parity operator. Next we study the spin
squeezing of the SCS $|\eta \rangle $ and the nonlinear SCS $|\eta ,t\rangle 
$.

\section{Spin Squeezing}

We first give a proof that there is no spin squeezing in the SCS along the $%
x,y,$ and $z$ directions, and then study the spin squeezing in the nonlinear
SCS.

\subsection{The SCS}

In order to calculate the squeezing parameter (\ref{eq:xi}) we need to know
the expectation values $\langle {\cal N}^k\rangle $ and $\langle
J_{-}^k\rangle (k=1,2).$ It is convenient to calculate $\langle {\cal N}%
^k\rangle $ by the generation function method. The generation function of
the SCS is given by 
\begin{equation}
G(\lambda )=\langle \eta |\lambda ^{{\cal N}}|\eta \rangle ={\frac{%
(1+\lambda |\eta |^2)^{2j}}{(1+|\eta |^2)^{2j}}},
\end{equation}
from which the factorial moments follow 
\begin{equation}
F(k)={\frac{d^kG(\lambda )}{d^k\lambda }}|_{\lambda =1}={\frac{|\eta
|^{2k}(2j)!}{(1+|\eta |^2)^k(2j-k)!}}.
\end{equation}
The factorial moments immediately give the expectation values of the
operators ${\cal N}$ and ${\cal N}^2$ and the variance of ${\cal N}$ 
\begin{eqnarray}
\langle {\cal N}\rangle  &=&F(1)={\frac{2j|\eta |^2}{1+|\eta |^2}},
\label{eq:n} \\
\langle {\cal N}^2\rangle  &=&F(2)+F(1)={\frac{2j|\eta |^2+4j^2|\eta |^4}{%
(1+|\eta |^2)^2}}  \label{eq:n2} \\
(\Delta {\cal N})^2 &=&{\frac{2j|\eta |^2}{(1+|\eta |^2)^2}.}  \label{eq:nnn}
\end{eqnarray}

From Eq.(\ref{eq:scs}) the expectation value $\langle J_{-}^k\rangle $ are
obtained as

\begin{equation}
\langle J_{-}^k\rangle ={\frac{\eta ^k(2j)!}{(1+|\eta |^2)^k(2j-k)!}}
\label{eq:jjjj}
\end{equation}
Now we calculate the squeezing parameter $\xi _z^2,$ which is rewritten as

\begin{equation}
\xi _z^2={\frac{2j(\Delta {\cal N})^2}{|\langle J_{-}\rangle |^2}}.
\label{eq:xixi}
\end{equation}
Then substituting Eqs.(\ref{eq:nnn}) and (\ref{eq:jjjj}) into Eq.(\ref
{eq:xixi}), we immediately obtain $\xi _z^2=1.$

To calculate $\xi _x^2$ and $\xi _y^2$ we need the identities

\begin{eqnarray}
J_x^2 &=&{\frac 14}[2j(2{\cal N}+1)-2{\cal N}^2+J_{+}^2+J_{-}^2],
\label{eq:xx} \\
J_y^2 &=&{\frac 14}[2j(2{\cal N}+1)-2{\cal N}^2-J_{+}^2-J_{-}^2],
\label{eq:yy}
\end{eqnarray}
which gives the expectation values of $J_x^2$ and $J_y^2$ in terms of $%
\langle {\cal N}\rangle ,\langle {\cal N}^2\rangle $ and  $\langle J_{\pm
}^2\rangle .$

From Eq.(\ref{eq:jjjj}) and the relation $J_z={\cal N}-j,$ we get

\begin{equation}
\langle J_x\rangle =\frac{j(\eta +\eta ^{*})}{1+|\eta |^2},\text{ }\langle
J_y\rangle =\frac{j(\eta ^{*}-\eta )}{i(1+|\eta |^2)},\text{ }\langle
J_z\rangle =\frac{j(|\eta |^2-1)}{1+|\eta |^2}  \label{eq:j}
\end{equation}
From Eqs.(\ref{eq:n}-\ref{eq:n2}), (\ref{eq:jjjj}), and (\ref{eq:xx}-\ref
{eq:j}), the variances of $J_x$ and $J_y$ are expressed as

\begin{eqnarray}
\langle \Delta J_x\rangle ^2 &=&\frac{j(1+|\eta |^4-\eta ^2-\eta ^{*2})}{%
2(1+|\eta |^2)^2}=\frac{\langle J_y\rangle ^2+\langle J_z\rangle ^2}{2j}, \\
\langle \Delta J_y\rangle ^2 &=&\frac{j(1+|\eta |^4+\eta ^2+\eta ^{*2})}{%
2(1+|\eta |^2)^2}=\frac{\langle J_z\rangle ^2+\langle J_x\rangle ^2}{2j}.
\end{eqnarray}
The above two equations directly lead to $\xi _x^2=\xi _y^2=1.$ Thus we have
shown that the squeezing parameters $\xi _x^2=\xi _y^2=\xi _z^2=1$ for the
SCS. That is to say, the SCS exhibits no squeezing in the $x$, $y$ and $z$
direction, irrespective of the complex $\eta $. We expect that the spin
squeezing exists in the nonlinear SCS.

\subsection{The nonlinear SCS}

We examine the spin squeezing in the nonlinear SCS $|\eta ,t\rangle .$ The
expectation values $\langle $ ${\cal N}\rangle $ ,$\langle {\cal N}^2\rangle 
$ and the variance $(\Delta {\cal N})^2$ are time independent and given by
Eqs.(\ref{eq:n}), (\ref{eq:n2}), and (\ref{eq:nnn}), respectively.

From Eq.(\ref{eq:nl}), we obtain the expectation value of $J_{-}^k$ on the
state $|\eta ,t\rangle $ as 
\begin{eqnarray}
\langle J_{-}^k\rangle  &=&\eta ^k(1+|\eta |^2)^{-2j}{\frac{(2j)!}{(2j-k)!}}
\nonumber \\
&&\times \sum_{n=0}^{2j-k}{{%
{2j-k \choose n}%
}}|\eta |^{2n}e^{it[F(n)-F(n+k)]}.  \label{eq:jminusk}
\end{eqnarray}
Of course Eq.(\ref{eq:jminusk}) reduces to Eq.(\ref{eq:jjjj}) at $t=0.$

By substituting Eqs.(\ref{eq:nnn}) and (\ref{eq:jminusk}) into (\ref{eq:xixi}),
the squeezing parameter $\xi _z^2$ is given by

\begin{equation}
\xi _z^2=\frac{{1}}{\left| \frac 1{(1+|\eta |^2)^{2j-1}}\sum_{n=0}^{2j-1}{{%
{2j-1  \choose n}%
}}|\eta |^{2n}e^{it[F(n)-F(n+1)]}\right| ^2}  \label{eq:xiz}
\end{equation}
For $M+1$ complex quantities $c_i(i=0...M),$ there is an inequality

\begin{equation}
|c_0+c_1+...+c_M|\leq |c_0|+|c_1|+...+|c_M|
\end{equation}
Using this inequality in Eq.(\ref{eq:xiz}), we find that

\begin{equation}
\xi _z^2\geq 1.
\end{equation}
So a general conclusion is made that no squeezing occurs in the $z$
direction for arbitrary nonlinear function $F({\cal N}).$ However spin
squeezing may exist in the $x$ or $y$ directions. Next we make numerical
calculations to show the spin squeezing.

\begin{figure}
\epsfig{width=8cm,file=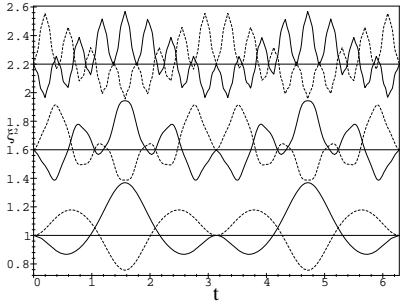}
\caption[]{
Squeezing parameters $\xi_\alpha^2 (\alpha=x,y)$ as a function of time $t$. We plot $\xi_\alpha^2$ for the nonlinear Hamtonian $F({\cal N})={\cal N}^2$, $\xi_\alpha^3+0.6$ for  $F({\cal N})={\cal N}^3$, and $\xi_\alpha^2+1.2$ for $F({\cal N})={\cal N}^4$. Solid line and dashed line correspond to $\xi_x^2$ and $\xi_y^2$, respectively. The parameters $\eta=0.1$ and $j=5$.} 
\end{figure}

Fig.1 gives the squeezing parameters $\xi _\alpha ^2(\alpha =x,y)$ as a
function of time $t$ for different nonlinear Hamiltonians $F({\cal N})={\cal %
N}^k(k=2,3,4)$. For small time $t$ we observe that the state is squeezed in
the $x$ direction other than the $y$ direction. As  $k$  increases, the
frequency of occurrence of spin squeezing increases. In most of the time we
also see that the spin squeezing alternatively appears in the $x$ and $y$
directions, i.e., when the state is squeezed in the $x(y)$ direction, it is
not squeezed in the $y(x)$ direction. The state can show no spin squeezing
in both the $x$ and $y$ directions in some small ranges of $t$, but it can
not show spin squeezing at the same time in the two directions.

\begin{figure}
\epsfig{width=8cm,file=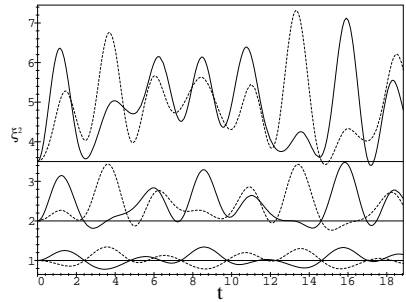}
\caption[]{Squeezing parameters $\xi_\alpha^2 (\alpha=x,y)$ as a function of time $t$ for the nonlinear Hamtonian $F({\cal N})=\sin(2{\cal N})$. Solid line and dashed line correspond to $\xi_x^2$ and $\xi_y^2$, respectively. We plot $\xi_\alpha^2$ for $\eta=0.1$, $\xi_\alpha^2+1$ for $\eta=0.2$, and $\xi_\alpha^2+2.5$ for $\eta=0.3$.The parameter $j=5$.} 
\end{figure}

It is interesting to consider the squeezing in the nonlinear Hamiltonian $H=\sin (a%
{\cal N})$ which can be realized in physical systems \cite{WMS}. The
numerical results are shown in Fig.2. For $\eta=0.1$ we observe that the spin
squeezing in the $x$ and $y$ directions appears alternatively  in the
beginning of the time evolution. For small time $t$, the state is squeezed
in the $y$ direction other than $x$ direction in contrast to Fig.1. We also observe that
the time range of spin squeezing decreases as the parameter $\eta$ increases, i.e., the squeezing does not
occur in most of the time as $\eta$ is large.

\section{Conclusions}

In conclusion we have given the definition and proposed an example of the
nonlinear SCS. We have studied the spin squeezing in both the SCS and the
nonlinear SCS. The main results are as follows:

1. The squeezing parameters $\xi _x^2=\xi _y^2=\xi _z^2=1$ for the SCS. That
is to say, the SCS is not squeezed in the $x$, $y$ and $z$ directions,
irrespective of the parameter $\eta $.

2. The nonlinear SCS shows no spin squeezing in the $z$ direction for
arbitrary nonlinear Hamiltonian $F({\cal N)}$.

3. The nonlinear SCS may be squeezed in the $x$ and $y$ directions. In most
of the time the squeezing appears alternatively in the $x$ and $y$
directions as time goes on. Also we observe that the state can not be
squeezed at the same time in the two directions.

The spin squeezing originates from the nonlinearity of the nonlinear SCS.
Then we expect that the spin squeezing exists in other nonlinear SCS with
difference nonlinear functions.

\acknowledgments
The author thanks for many helpful discussions with Klaus M\o lmer, Anders S%
\o rensen, and Bin Shu. This work is supported by the Information Society
Technologies Programme IST-1999-11053, EQUIP, action line 6-2-1.

\end{document}